# Search for the rare decays $D \to h(h^{(\prime)})e^+e^-$

M. Ablikim,[1] M. N. Achasov,[10,d] S. Ahmed,[15] M. Albrecht,[4] M. Alekseev,[56a,56c] A. Amoroso,[56a,56c] F. F. An,[1] Q. An,[53,43] J. Z. Bai,[1] Y. Bai,[42] O. Bakina,[27] R. Baldini Ferroli,[23a] Y. Ban,[35] K. Begzsuren,[25] D. W. Bennett,[22] J. V. Bennett,[5] N. Berger,[26] M. Bertani,[23a] D. Bettoni,[24a] F. Bianchi,[56a,56c] E. Boger,[27,b] I. Boyko,[27] R. A. Briere,[5] H. Cai,[58] X. Cai,[1,43] O. Cakir,[46a] A. Calcaterra,[23a] G. F. Cao,[1,47] S. A. Cetin,[46b] J. Chai,[56c] J. F. Chang,[1,43] G. Chelkov,[27,b,c] G. Chen,[1] H. S. Chen,[1,47] J. C. Chen,[1] M. L. Chen,[1,43] P. L. Chen,[54] S. J. Chen,[33] X. R. Chen,[30] Y. B. Chen,[1,43] W. Cheng,[56c] X. K. Chu,[35] G. Cibinetto,[24a] F. Cossio,[56c] H. L. Dai,[1,43] J. P. Dai,[38,h] A. Dbeyssi,[15] D. Dedovich,[27] Z. Y. Deng,[1] A. Denig,[26] I. Denysenko,[27] M. Destefanis,[56a,56c] F. De Mori,[56a,56c] Y. Ding,[31] C. Dong,[34] J. Dong,[1,43] L. Y. Dong,[1,47] M. Y. Dong,[1,43,47] Z. L. Dou,[33] S. X. Du,[61] P. F. Duan,[1] J. Fang,[1,43] S. S. Fang,[1,47] Y. Fang,[1] R. Farinelli,[24a,24b] L. Fava,[56b,56c] S. Fegan,[26] F. Feldbauer,[4] G. Felici,[23a] C. Q. Feng,[53,43] E. Fioravanti,[24a] M. Fritsch,[4] C. D. Fu,[1] Q. Gao,[1] X. L. Gao,[53,43] Y. Gao,[45] Y. G. Gao,[6] Z. Gao,[53,43] B. Garillon,[26] I. Garzia,[24a] A. Gilman,[50] K. Goetzen,[11] L. Gong,[34] W. X. Gong,[1,43] W. Gradl,[26] M. Greco,[56a,56c] M. H. Gu,[1,43] Y. T. Gu,[13] A. Q. Guo,[1] R. P. Guo,[1,47] Y. P. Guo,[26] A. Guskov,[27] Z. Haddadi,[29] S. Han,[58] X. Q. Hao,[16] F. A. Harris,[48] K. L. He,[1,47] X. Q. He,[52] F. H. Heinsius,[4] T. Held,[4] Y. K. Heng,[1,43,47] Z. L. Hou,[1] H. M. Hu,[1,47] J. F. Hu,[38,h] T. Hu,[1,43,47] Y. Hu,[1] G. S. Huang,[53,43] J. S. Huang,[16] X. T. Huang,[37] X. Z. Huang,[33] Z. L. Huang,[31] T. Hussain,[55] W. Ikegami Andersson,[57] M. Irshad,[53,43] Q. Ji,[1] Q. P. Ji,[16] X. B. Ji,[1,47] X. L. Ji,[1,43] X. S. Jiang,[1,43,47] X. Y. Jiang,[34] J. B. Jiao,[37] Z. Jiao,[18] D. P. Jin,[1,43,47] S. Jin,[1,47] Y. Jin,[49] T. Johansson,[57] A. Julin,[50] N. Kalantar-Nayestanaki,[29] X. S. Kang,[34] M. Kavatsyuk,[29] B. C. Ke,[1] I. K. Keshk,[4] T. Khan,[53,43] A. Khoukaz,[51] P. Kiese,[26] R. Kiuchi,[1] R. Kliemt,[11] L. Koch,[28] O. B. Kolcu,[46b,f] B. Kopf,[4] M. Kornicer,[48] M. Kuemmel,[4] M. Kuessner,[4] A. Kupsc,[57] M. Kurth,[1] W. Kühn,[28] J. S. Lange,[28] P. Larin,[15] L. Lavezzi,[56c] H. Leithoff,[26] C. Li,[57] Cheng Li,[53,43] D. M. Li,[61] F. Li,[1,43] F. Y. Li,[35] G. Li,[1] H. B. Li,[1,47] H. J. Li,[1,47] J. C. Li,[1] J. W. Li,[41] Jin Li,[36] K. J. Li,[44] Kang Li,[14] Ke Li,[1] Lei Li,[3] P. L. Li,[53,43] P. R. Li,[47,7] Q. Y. Li,[37] W. D. Li,[1,47] W. G. Li,[1] X. L. Li,[37] X. N. Li,[1,43] X. Q. Li,[34] Z. B. Li,[44] H. Liang,[53,43] Y. F. Liang,[40] Y. T. Liang,[28] G. R. Liao,[12] L. Z. Liao,[1,47] J. Libby,[21] C. X. Lin,[44] D. X. Lin,[15] B. Liu,[38,h] B. J. Liu,[1] C. X. Liu,[1] D. Liu,[53,43] D. Y. Liu,[38,h] F. H. Liu,[39] Fang Liu,[1] Feng Liu,[6] H. B. Liu,[13] H. L. Liu,[42] H. M. Liu,[1,47] Huanhuan Liu,[1] Huihui Liu,[17] J. B. Liu,[53,43] J. Y. Liu,[1,47] K. Liu,[45] K. Y. Liu,[31] Ke Liu,[6] L. D. Liu,[35] Q. Liu,[47] S. B. Liu,[53,43] X. Liu,[30] Y. B. Liu,[34] Z. A. Liu,[1,43,47] Zhiqing Liu,[26] Y. F. Long,[35] X. C. Lou,[1,43,47] H. J. Lu,[18] J. G. Lu,[1,43] Y. Lu,[1] Y. P. Lu,[1,43] C. L. Luo,[32] M. X. Luo,[60] T. Luo,[9,j] X. L. Luo,[1,43] S. Lusso,[56c] X. R. Lyu,[47] F. C. Ma,[31] H. L. Ma,[1] L. L. Ma,[37] M. M. Ma,[1,47] Q. M. Ma,[1] T. Ma,[1] X. N. Ma,[34] X. Y. Ma,[1,43] Y. M. Ma,[37] F. E. Maas,[15] M. Maggiora,[56a,56c] S. Maldaner,[26] Q. A. Malik,[55] A. Mangoni,[23b] Y. J. Mao,[35] Z. P. Mao,[1] S. Marcello,[56a,56c] Z. X. Meng,[49] J. G. Messchendorp,[29] G. Mezzadri,[24b] J. Min,[1,43] R. E. Mitchell,[22] X. H. Mo,[1,43,47] Y. J. Mo,[6] C. Morales Morales,[15] N. Yu. Muchnoi,[10,d] H. Muramatsu,[50] A. Mustafa,[4] Y. Nefedov,[27] F. Nerling,[11] I. B. Nikolaev,[10,d] Z. Ning,[1,43] S. Nisar,[8] S. L. Niu,[1,43] X. Y. Niu,[1,47] S. L. Olsen,[36,k] Q. Ouyang,[1,43,47] S. Pacetti,[23b] Y. Pan,[53,43] M. Papenbrock,[57] P. Patteri,[23a] M. Pelizaeus,[4] J. Pellegrino,[56a,56c] H. P. Peng,[53,43] Z. Y. Peng,[13] K. Peters,[11,g] J. Pettersson,[57] J. L. Ping,[32] R. G. Ping,[1,47] A. Pitka,[4] R. Poling,[50] V. Prasad,[53,43] H. R. Qi,[2] M. Qi,[33] T. Y. Qi,[2] S. Qian,[1,43] C. F. Qiao,[47] N. Qin,[58] X. S. Qin,[4] Z. H. Qin,[1,43] J. F. Qiu,[1] S. Q. Qu,[34] K. H. Rashid,[55,i] C. F. Redmer,[26] M. Richter,[4] M. Ripka,[26] A. Rivetti,[56c] M. Rolo,[56c] G. Rong,[1,47] Ch. Rosner,[15] A. Sarantsev,[27,e] M. Savrié,[24b] K. Schoenning,[57] W. Shan,[19] X. Y. Shan,[53,43] M. Shao,[53,43] C. P. Shen,[2] P. X. Shen,[34] X. Y. Shen,[1,47] H. Y. Sheng,[1] X. Shi,[1,43] J. J. Song,[37] W. M. Song,[37] X. Y. Song,[1] S. Sosio,[56a,56c] C. Sowa,[4] S. Spataro,[56a,56c] G. X. Sun,[1] J. F. Sun,[16] L. Sun,[58] S. S. Sun,[1,47] X. H. Sun,[1] Y. J. Sun,[53,43] Y. K. Sun,[53,43] Y. Z. Sun,[1] Z. J. Sun,[1,43] Z. T. Sun,[1] Y. T. Tan,[53,43] C. J. Tang,[40] G. Y. Tang,[1] X. Tang,[1] I. Tapan,[46c] M. Tiemens,[29] B. Tsednee,[25] I. Uman,[46d] B. Wang,[1] B. L. Wang,[47] D. Wang,[35] D. Y. Wang,[35] Dan Wang,[47] K. Wang,[1,43] L. L. Wang,[1] L. S. Wang,[1] M. Wang,[37] Meng Wang,[1,47] P. Wang,[1] P. L. Wang,[1] W. P. Wang,[53,43] X. F. Wang,[45] Y. Wang,[53,43] Y. F. Wang,[1,43,47] Z. Wang,[1,43] Z. G. Wang,[1,43] Z. Y. Wang,[1] Zongyuan Wang,[1,47] T. Weber,[4] D. H. Wei,[12] P. Weidenkaff,[26] S. P. Wen,[1] U. Wiedner,[4] M. Wolke,[57] L. H. Wu,[1] L. J. Wu,[1,47] Z. Wu,[1,43] L. Xia,[53,43] Y. Xia,[20] D. Xiao,[1] Y. J. Xiao,[1,47] Z. J. Xiao,[32] Y. G. Xie,[1,43] Y. H. Xie,[6] X. A. Xiong,[1,47] Q. L. Xiu,[1,43] G. F. Xu,[1] J. J. Xu,[1,47] L. Xu,[1] Q. J. Xu,[14] Q. N. Xu,[47] X. P. Xu,[41] F. Yan,[54] L. Yan,[56a,56c] W. B. Yan,[53,43] W. C. Yan,[2] Y. H. Yan,[20] H. J. Yang,[38,h] H. X. Yang,[1] L. Yang,[58] R. X. Yang,[53,43] Y. H. Yang,[33] Y. X. Yang,[12] Yifan Yang,[1,47] Z. Q. Yang,[20] M. Ye,[1,43] M. H. Ye,[7] J. H. Yin,[1] Z. Y. You,[44] B. X. Yu,[1,43,47] C. X. Yu,[34] J. S. Yu,[20] J. S. Yu,[30] C. Z. Yuan,[1,47] Y. Yuan,[1] A. Yuncu,[46b,a] A. A. Zafar,[55] Y. Zeng,[20] B. X. Zhang,[1] B. Y. Zhang,[1,43] C. C. Zhang,[1] D. H. Zhang,[1] H. H. Zhang,[44] H. Y. Zhang,[1,43] J. Zhang,[1,47] J. L. Zhang,[59] J. Q. Zhang,[4] J. W. Zhang,[1,43,47] J. Y. Zhang,[1] J. Z. Zhang,[1,47] K. Zhang,[1,47] L. Zhang,[45] T. J. Zhang,[38,h] X. Y. Zhang,[37] Y. Zhang,[53,43] Y. H. Zhang,[1,43] Y. T. Zhang,[53,43] Yang Zhang,[1] Yao Zhang,[1] Yu Zhang,[47] Z. H. Zhang,[6] Z. P. Zhang,[53] Z. Y. Zhang,[58] G. Zhao,[1] J. W. Zhao,[1,43] J. Y. Zhao,[1,47] J. Z. Zhao,[1,43] Lei Zhao,[53,43] Ling Zhao,[1] M. G. Zhao,[34] Q. Zhao,[1] S. J. Zhao,[61] T. C. Zhao,[1] Y. B. Zhao,[1,43] Z. G. Zhao,[53,43] A. Zhemchugov,[27,b] B. Zheng,[54] J. P. Zheng,[1,43] W. J. Zheng,[37] Y. H. Zheng,[47] B. Zhong,[32] L. Zhou,[1,43] Q. Zhou,[1,47] X. Zhou,[58] X. K. Zhou,[53,43] X. R. Zhou,[53,43]

X. Y. Zhou,[1] Xiaoyu Zhou,[20] Xu Zhou,[20] A. N. Zhu,[1,47] J. Zhu,[34] J. Zhu,[44] K. Zhu,[1] K. J. Zhu,[1,43,47] S. Zhu,[1] S. H. Zhu,[52] X. L. Zhu,[45] Y. C. Zhu,[53,43] Y. S. Zhu,[1,47] Z. A. Zhu,[1,47] J. Zhuang,[1,43] B. S. Zou,[1] and J. H. Zou[1]

(BESIII Collaboration)

[1]*Institute of High Energy Physics, Beijing 100049, People's Republic of China*
[2]*Beihang University, Beijing 100191, People's Republic of China*
[3]*Beijing Institute of Petrochemical Technology, Beijing 102617, People's Republic of China*
[4]*Bochum Ruhr-University, D-44780 Bochum, Germany*
[5]*Carnegie Mellon University, Pittsburgh, Pennsylvania 15213, USA*
[6]*Central China Normal University, Wuhan 430079, People's Republic of China*
[7]*China Center of Advanced Science and Technology, Beijing 100190, People's Republic of China*
[8]*COMSATS Institute of Information Technology, Lahore, Defence Road, Off Raiwind Road, 54000 Lahore, Pakistan*
[9]*Fudan University, Shanghai 200443, People's Republic of China*
[10]*G.I. Budker Institute of Nuclear Physics SB RAS (BINP), Novosibirsk 630090, Russia*
[11]*GSI Helmholtzcentre for Heavy Ion Research GmbH, D-64291 Darmstadt, Germany*
[12]*Guangxi Normal University, Guilin 541004, People's Republic of China*
[13]*Guangxi University, Nanning 530004, People's Republic of China*
[14]*Hangzhou Normal University, Hangzhou 310036, People's Republic of China*
[15]*Helmholtz Institute Mainz, Johann-Joachim-Becher-Weg 45, D-55099 Mainz, Germany*
[16]*Henan Normal University, Xinxiang 453007, People's Republic of China*
[17]*Henan University of Science and Technology, Luoyang 471003, People's Republic of China*
[18]*Huangshan College, Huangshan 245000, People's Republic of China*
[19]*Hunan Normal University, Changsha 410081, People's Republic of China*
[20]*Hunan University, Changsha 410082, People's Republic of China*
[21]*Indian Institute of Technology Madras, Chennai 600036, India*
[22]*Indiana University, Bloomington, Indiana 47405, USA*
[23a]*INFN Laboratori Nazionali di Frascati, I-00044, Frascati, Italy*
[23b]*INFN and University of Perugia, I-06100, Perugia, Italy*
[24a]*INFN Sezione di Ferrara, I-44122, Ferrara, Italy*
[24b]*University of Ferrara, I-44122, Ferrara, Italy*
[25]*Institute of Physics and Technology, Peace Ave. 54B, Ulaanbaatar 13330, Mongolia*
[26]*Johannes Gutenberg University of Mainz, Johann-Joachim-Becher-Weg 45, D-55099 Mainz, Germany*
[27]*Joint Institute for Nuclear Research, 141980 Dubna, Moscow region, Russia*
[28]*Justus-Liebig-Universitaet Giessen, II. Physikalisches Institut, Heinrich-Buff-Ring 16, D-35392 Giessen, Germany*
[29]*KVI-CART, University of Groningen, NL-9747 AA Groningen, The Netherlands*
[30]*Lanzhou University, Lanzhou 730000, People's Republic of China*
[31]*Liaoning University, Shenyang 110036, People's Republic of China*
[32]*Nanjing Normal University, Nanjing 210023, People's Republic of China*
[33]*Nanjing University, Nanjing 210093, People's Republic of China*
[34]*Nankai University, Tianjin 300071, People's Republic of China*
[35]*Peking University, Beijing 100871, People's Republic of China*
[36]*Seoul National University, Seoul 151-747, Korea*
[37]*Shandong University, Jinan 250100, People's Republic of China*
[38]*Shanghai Jiao Tong University, Shanghai 200240, People's Republic of China*
[39]*Shanxi University, Taiyuan 030006, People's Republic of China*
[40]*Sichuan University, Chengdu 610064, People's Republic of China*
[41]*Soochow University, Suzhou 215006, People's Republic of China*
[42]*Southeast University, Nanjing 211100, People's Republic of China*
[43]*State Key Laboratory of Particle Detection and Electronics, Beijing 100049, Hefei 230026, People's Republic of China*
[44]*Sun Yat-Sen University, Guangzhou 510275, People's Republic of China*
[45]*Tsinghua University, Beijing 100084, People's Republic of China*
[46a]*Ankara University, 06100 Tandogan, Ankara, Turkey*
[46b]*Istanbul Bilgi University, 34060 Eyup, Istanbul, Turkey*
[46c]*Uludag University, 16059 Bursa, Turkey*
[46d]*Near East University, Nicosia, North Cyprus, Mersin 10, Turkey*
[47]*University of Chinese Academy of Sciences, Beijing 100049, People's Republic of China*

[48]University of Hawaii, Honolulu, Hawaii 96822, USA
[49]University of Jinan, Jinan 250022, People's Republic of China
[50]University of Minnesota, Minneapolis, Minnesota 55455, USA
[51]University of Muenster, Wilhelm-Klemm-Str. 9, 48149 Muenster, Germany
[52]University of Science and Technology Liaoning, Anshan 114051, People's Republic of China
[53]University of Science and Technology of China, Hefei 230026, People's Republic of China
[54]University of South China, Hengyang 421001, People's Republic of China
[55]University of the Punjab, Lahore-54590, Pakistan
[56a]University of Turin, I-10125 Turin, Italy
[56b]University of Eastern Piedmont, I-15121 Alessandria, Italy
[56c]INFN, I-10125 Turin, Italy
[57]Uppsala University, Box 516, SE-75120 Uppsala, Sweden
[58]Wuhan University, Wuhan 430072, People's Republic of China
[59]Xinyang Normal University, Xinyang 464000, People's Republic of China
[60]Zhejiang University, Hangzhou 310027, People's Republic of China
[61]Zhengzhou University, Zhengzhou 450001, People's Republic of China

We search for rare decays of $D$ mesons to hadrons accompanied by an electron-positron pair ($h(h^{(\prime)})e^+e^-$), using an $e^+e^-$ collision sample corresponding to an integrated luminosity of 2.93 fb$^{-1}$ collected with the BESIII detector at $\sqrt{s} = 3.773$ GeV. No significant signals are observed, and the corresponding upper limits on the branching fractions at the 90% confidence level are determined. The sensitivities of the results are at the level of $10^{-5}$–$10^{-6}$, providing a large improvement over previous searches.

## I. INTRODUCTION

In the standard model (SM), the decay of a $D$ meson into hadrons accompanied by a lepton pair proceeds via the quark process $c \to ul^+l^-$ ($l = e$ or $\mu$). This is known as a flavor changing neutral current (FCNC) process, which is forbidden at tree level in the SM. It can happen only through a loop diagram because of the suppression of the Glashow-Iliopoulos-Maiani (GIM) mechanism [1], leading to a very small branching fraction (BF) theoretically, which would not exceed the level of $10^{-9}$ [2–4]. Compared to similar FCNC processes in $B$- and $K$-meson decays, the GIM suppression in FCNC decays of the $D$ meson is much stronger, as better diagram cancellation occurs due to the down-type quarks involved. However, possible new physics (NP) beyond the SM can significantly increase the decay rates of these short distance (SD) processes. Hence, they can serve as clean channels in experiments to search for NP [2,3].

However, these $D$-meson-decay rates are also contributed by long distance (LD) effects through (virtual) vector meson ($V^{(\star)}$) decays, like $D \to hV^{(\star)}, V^{(\star)} \to l^+l^-$, even above the level of $10^{-6}$ [3,4]. Therefore, FCNC processes are potentially overshadowed by LD effects. In such case, a measurement of the angular dependence or $CP$ asymmetry is required to figure out the SD effects and to test the SM prediction.

In recent years, the four-body decays of $D^0$ mesons with a $\mu^+\mu^-$ pair in final state, i.e., $D^0 \to K^-\pi^+\mu^+\mu^-$, $K^-K^+\mu^+\mu^-$ and $\pi^-\pi^+\mu^+\mu^-$, have been observed at LHCb [5,6], with the decay rates at the level of $10^{-7}$, indicating significant LD contributions. However, no evidence for the $e^+e^-$ modes has yet been reported.

[a]Also at Bogazici University, 34342 Istanbul, Turkey.
[b]Also at the Moscow Institute of Physics and Technology, Moscow 141700, Russia.
[c]Also at the Functional Electronics Laboratory, Tomsk State University, Tomsk, 634050, Russia.
[d]Also at the Novosibirsk State University, Novosibirsk, 630090, Russia.
[e]Also at the NRC "Kurchatov Institute", PNPI, 188300, Gatchina, Russia.
[f]Also at Istanbul Arel University, 34295 Istanbul, Turkey.
[g]Also at Goethe University Frankfurt, 60323 Frankfurt am Main, Germany.
[h]Also at Key Laboratory for Particle Physics, Astrophysics and Cosmology, Ministry of Education; Shanghai Key Laboratory for Particle Physics and Cosmology; Institute of Nuclear and Particle Physics, Shanghai 200240, People's Republic of China.
[i]Also at Government College Women University, Sialkot—51310. Punjab, Pakistan.
[j]Also at Key Laboratory of Nuclear Physics and Ion-beam Application (MOE) and Institute of Modern Physics, Fudan University, Shanghai 200443, People's Republic of China.
[k]Present address: Center for Underground Physics, Institute for Basic Science, Daejeon 34126, Korea.

The three-body and four-body decays of $D^0$ mesons involving $e^+e^-$ pairs were searched for by the CLEO and E791 Collaborations [7,8]. The current upper limits (ULs) on their branching fractions at the 90% confidence level (CL) are at the level of $10^{-4}$–$10^{-5}$. The analogous $D^+$ decays are less well studied, and only three-body decays have been searched for by the *BABAR* and LHCb Collaborations [9].

In this paper, using an $e^+e^-$ collision sample corresponding to an integrated luminosity of 2.93 $\mathrm{fb}^{-1}$ [10] collected with the BESIII detector at $\sqrt{s} = 3.773$ GeV, we perform a search for the rare decays of $D \to h(h^{(\prime)})e^+e^-$, where $h^{(\prime)}$ are hadrons. To avoid possible bias, a blind analysis is carried out based on Monte Carlo (MC) simulations to validate the analysis strategy, the results are opened only after the analysis strategy is fixed.

## II. THE BESIII DETECTOR AND MC SIMULATION

The Beijing Spectrometer (BESIII) detects $e^+e^-$ collisions in the double-ring collider BEPCII. BESIII is a general-purpose detector [11] with 93% coverage of the full solid angle. From the interaction point (IP) to the outside, BESIII is equipped with a main drift chamber (MDC) consisting of 43 layers of drift cells, a time-of-flight (TOF) counter with double-layer scintillator in the barrel part and single-layer scintillator in the end-cap part, an electromagnetic calorimeter (EMC) composed of 6240 CsI (Tl) crystals, a superconducting solenoid magnet providing a magnetic field of 1.0 T along the beam direction, and a muon counter containing multilayer resistive plate chambers installed in the steel flux-return yoke of the magnet. The MDC spatial resolution is 135 $\mu$m and the momentum resolution is 0.5% for a charged track with transverse momentum of 1 GeV/c. The energy resolution for a photon at 1 GeV in the EMC is 2.5% in the barrel region and 5.0% in the endcap region. More details of the spectrometer can be found in Ref. [11].

Monte Carlo simulation serves to estimate the detection efficiencies and to understand background contamination. High statistics MC samples are generated with a GEANT4-based [12] software package, which includes the descriptions of the geometry of the spectrometer and interactions of particles with the detector materials. KKMC [13] is used to model the beam energy spread and the initial-state radiation (ISR) in the $e^+e^-$ annihilations. The "inclusive" MC samples consist of the production of $D\bar{D}$ pairs with quantum coherence for all neutral $D$ modes, the non-$D\bar{D}$ decays of $\psi(3770)$, the ISR production of low mass $\psi$ states, and continuum processes. Known decays recorded by the Particle Data Group (PDG) [9] are simulated with EVTGEN [14] and the unknown decays with LUNDCHARM model [15]. The final-state radiation (FSR) of charged tracks is taken into account with the PHOTOS package [16]. The equivalent luminosity of the inclusive MC samples is about 10 times that of the data. The signal processes are generated using the phase space model (PHSP) of EVTGEN. For each signal channel, 200000 events are simulated.

## III. DATA ANALYSIS

Since the center-of-mass energy of 3.773 GeV is close to the $D\bar{D}$ mass threshold, the pair of $D^+D^-$ or $D^0\bar{D}^0$ mesons is produced nearly at rest without any additional hadrons. Hence, it is straightforward to use a double tagging approach [17] to measure absolute BFs, based on the following equation

$$\mathcal{B} = \frac{n_{\mathrm{sig,tag}}}{\sum_i n^i_{\mathrm{tag}} \cdot \frac{\varepsilon^i_{\mathrm{sig,tag}}}{\varepsilon^i_{\mathrm{tag}}}} = \frac{n_{\mathrm{sig,tag}}}{n_{\mathrm{tag}} \cdot \varepsilon_{\mathrm{sig}}}. \tag{1}$$

Here, $i$ denotes the different single-tag (ST) modes of hadronic decays, and $n^i_{\mathrm{tag}}$ is the yield of the $\bar{D}$ meson of ST tag mode $i$. $n_{\mathrm{sig,tag}}$ is the number of $D$ rare decay candidate events in which a ST $\bar{D}$ meson is detected, so called double-tag (DT) events. Finally, $\varepsilon^i_{\mathrm{tag}}$ and $\varepsilon^i_{\mathrm{sig,tag}}$ are the corresponding ST and DT detection efficiencies. The average signal efficiency over different ST modes can be calculated to be $\varepsilon_{\mathrm{sig}} = (\sum_i n^i_{\mathrm{tag}} \cdot \varepsilon^i_{\mathrm{tag,sig}}/\varepsilon^i_{\mathrm{tag}})/n_{\mathrm{tag}}$, where $n_{\mathrm{tag}}$ is the total number of ST events $n_{\mathrm{tag}} = \sum_i n^i_{\mathrm{tag}}$. Note that in this paper, charge conjugated modes are always implied.

### A. ST event selection and yields

The ST modes used to tag $D^-$ candidates are $K^+\pi^-\pi^-$, $K^+\pi^-\pi^-\pi^0$, $K^0_S\pi^-$, $K^0_S\pi^-\pi^0$, $K^0_S\pi^+\pi^-\pi^-$ and $K^+K^-\pi^-$, while the modes $K^+\pi^-, K^+\pi^-\pi^0$ and $K^+\pi^-\pi^+\pi^-$, with $\pi^0 \to \gamma\gamma$ and $K^0_S \to \pi^+\pi^-$, are used to tag $\bar{D}^0$. The sum of the BFs is about 27.7% for the six $D^-$ decays, and 26.7% for the three $\bar{D}^0$ decays. ST candidates are reconstructed from all possible combinations of final state particles, according to the following selection criteria.

Momenta and impact parameters of charged tracks are measured by the MDC. Charged tracks (except for those of $K^0_S$ decays) are required to satisfy $|\cos\theta| < 0.93$, where $\theta$ is the polar angle with respect to the beam axis, and have a distance of closest approach to the interaction point (IP) within $\pm 10$ cm along the beam direction and within $\pm 1$ cm in the plane perpendicular to the beam axis. Particle identification (PID) is implemented by combining the specific energy loss ($\mathrm{d}E/\mathrm{d}x$) in the MDC and the time of flight measured from the TOF to form PID confidence levels (CL) for each particle hypothesis. For a charged $\pi(K)$ candidate, the CL of the $\pi(K)$ hypothesis is required to be larger than that of the $K(\pi)$ hypothesis.

Photon candidates are reconstructed from clusters of deposited energy in the EMC. The energies of photon candidates must be larger than 25 MeV for $|\cos\theta| < 0.8$ (barrel) or 50 MeV for $0.86 < |\cos\theta| < 0.92$ (end cap). To suppress fake photons due to electronic noise or beam background, the shower time must be less than 700 ns from the event start time [18]. The photon candidates are required to be at least 20° away from any charged track.

The $\pi^0$ candidates are reconstructed from pairs of photons of which at least one is reconstructed in the barrel. The invariant mass of the photon pair, $M_{\gamma\gamma}$, is required to lie in the range $(0.115, 0.150)$ GeV/c$^2$. To improve the resolution, we further constrain the invariant mass of each photon pair to the nominal $\pi^0$ mass by a kinematic fit, and the updated four-momentum of the $\pi^0$ will be used in the further analysis.

The $K_S^0$ candidates are reconstructed via $K_S^0 \to \pi^+\pi^-$ using a vertex-constrained fit to all pairs of oppositely charged tracks, without PID requirements. The distance of closest approach of a charged track to the IP is required to be less than ±20 cm along the beam direction, without any requirement in the transverse plane. The $\chi^2$ of the vertex fit is required to be less than 100. The invariant mass of the $\pi^+\pi^-$ pair, $M_{\pi^+\pi^-}$, is required to be within $(0.487, 0.511)$ GeV/c$^2$, corresponding to three times the experimental mass resolution.

Two variables, the beam-constrained mass, $M_{\rm BC}^{\rm tag}$, and the energy difference, $\Delta E_{\rm tag}$, which are defined as

$$M_{\rm BC}^{\rm tag} \equiv \sqrt{E_{\rm beam}^2/c^4 - |\vec{p}_{\bar{D}}|^2/c^2},$$

$$\Delta E_{\rm tag} \equiv E_{\bar{D}} - E_{\rm beam},$$

are used to identify the tag candidates. Here, $\vec{p}_{\bar{D}}$ and $E_{\bar{D}}$ are the momentum and energy of the ST $\bar{D}$ candidate in the rest frame of the initial $e^+e^-$ system, and $E_{\rm beam}$ is the beam energy. Signal events peak around the nominal $\bar{D}$ mass in the $M_{\rm BC}^{\rm tag}$ distribution and around zero in the $\Delta E_{\rm tag}$ distribution. The boundaries of the $\Delta E_{\rm tag}$ requirements are determined from MC simulation, and set at approximately $(-3\sigma, 3\sigma)$ for the modes with only charged tracks in final state, and $(-4\sigma, +3.5\sigma)$ for those including a $\pi^0$ in the final state, due to the asymmetric $\Delta E$ distribution. Here, $\sigma$ is the standard deviation of $\Delta E_{\rm tag}$. In each event, only the combination with the smallest $|\Delta E_{\rm tag}|$ is kept for each ST mode.

After applying the $\Delta E_{\rm tag}$ requirements as listed in Table I for the different ST modes, the $M_{\rm BC}^{\rm tag}$ distributions are shown in Fig. 1. The corresponding ST yields are extracted by performing maximum likelihood fits to the $M_{\rm BC}^{\rm tag}$ distribution, where in each mode the signal is modeled with a MC-derived signal shape convolved with a smearing Gaussian function representing the resolution difference between data and MC simulation, and the backgrounds are modeled with an ARGUS function [19]. Based on the fit results, the total ST yields found in data are summarized in Table I together with the MC-determined detection efficiencies.

Events with a ST candidate fulfilling the additional requirement of $M_{\rm BC}^{\rm tag}$ to be within $(1.863, 1.879)$ GeV/c$^2$ for charged $D$ modes and $(1.858, 1.874)$ GeV/c$^2$ for neutral $D$ modes are used to search for signal candidates as described in the following.

TABLE I. Requirements on $\Delta E_{\rm tag}$, MC-determined detection efficiencies $\varepsilon_{\rm tag}^{\rm i,MC}$ and signal yields for the different ST modes.

| $D^-$ decays | $\Delta E_{\rm tag}$ (GeV) | $\varepsilon_{\rm tag}^{\rm i,MC}$ (%) | $n_{\rm tag}^i$ |
|---|---|---|---|
| $K^+\pi^-\pi^-$ | $(-0.022, 0.021)$ | $50.47 \pm 0.06$ | $755661 \pm 922$ |
| $K^+\pi^-\pi^-\pi^0$ | $(-0.060, 0.034)$ | $24.65 \pm 0.05$ | $231322 \pm 729$ |
| $K_S^0\pi^-$ | $(-0.019, 0.021)$ | $54.44 \pm 0.17$ | $95346 \pm 330$ |
| $K_S^0\pi^-\pi^0$ | $(-0.071, 0.041)$ | $27.44 \pm 0.06$ | $210535 \pm 638$ |
| $K_S^0\pi^+\pi^-\pi^-$ | $(-0.025, 0.023)$ | $31.80 \pm 0.09$ | $119249 \pm 451$ |
| $K^+K^-\pi^-$ | $(-0.019, 0.018)$ | $40.71 \pm 0.16$ | $64904 \pm 259$ |

| $\bar{D}^0$ decays | $\Delta E_{\rm tag}$ (GeV) | $\varepsilon_{\rm tag}^{\rm i,MC}$ (%) | $n_{\rm tag}^i$ |
|---|---|---|---|
| $K^+\pi^-$ | $(-0.023, 0.022)$ | $64.64 \pm 0.03$ | $523265 \pm 763$ |
| $K^+\pi^-\pi^0$ | $(-0.064, 0.035)$ | $33.60 \pm 0.01$ | $1022697 \pm 1448$ |
| $K^+\pi^-\pi^+\pi^-$ | $(-0.026, 0.023)$ | $38.26 \pm 0.02$ | $707936 \pm 1129$ |

### B. Signal event selection and yields

Signal candidates of $D^+$ decaying to $\pi^+\pi^0e^+e^-$, $K^+\pi^0e^+e^-$, $K_S^0\pi^+e^+e^-$, and $K_S^0K^+e^+e^-$, and $D^0$ decaying to $K^-K^+e^+e^-$, $\pi^+\pi^-e^+e^-$, $K^-\pi^+e^+e^-$, $\pi^0e^+e^-$, $\eta e^+e^-$, $\omega e^+e^-$, and $K_S^0e^+e^-$ are searched for in the remaining charged tracks and showers recoiling against the ST $\bar{D}$ mesons. The selection criteria for the charged tracks and neutral showers are the same as those used in the ST event selection. Positrons and electrons are distinguished from other charged particles by combining the d$E$/d$x$, TOF and EMC information. The determined particle identification CL, $\mathcal{L}$, is required to satisfy $\mathcal{L}(e) > 0$ and $\mathcal{L}(e)/(\mathcal{L}(e) + \mathcal{L}(\pi) + \mathcal{L}(K)) > 0.8$. Furthermore, the energy deposited in the EMC divided by the momentum measured in the MDC, $E/pc$, is required to be larger than 0.8 for either the positron or electron. By studying the inclusive MC samples, we find that the selected $e^+e^-$ pairs dominantly originate from $\gamma$-conversion events, where the photons are from the decay of intermediate states. To suppress these backgrounds, the vertex of the $e^+e^-$ pair is reconstructed [20,21], and the distance from the IP to the reconstructed vertex in the $x - y$ plane $R_{xy}$ is required to be out of range (2.0, 8.0) cm, where the $\gamma$-conversion occurs.

To veto the contribution from $D \to h(h^{(\prime)})\phi$, $\phi \to e^+e^-$, the $e^+e^-$ invariant mass $M_{e^+e^-}$ is required to be outside of the $\phi$ mass region, defined as $(0.935, 1.053)$ GeV/c$^2$.

An $\eta$ candidate is reconstructed via its $\gamma\gamma$ decay mode by requiring $M_{\gamma\gamma}$ within $(0.505, 0.570)$ GeV/c$^2$. A kinematic fit constraining $M_{\gamma\gamma}$ to the nominal $\eta$ mass is applied, and the candidate with the smallest $\chi^2$ is kept under the requirement $\chi^2 < 20$. Similarly, candidate $\pi^0$ decaying into $\gamma\gamma$ are selected by requiring $M_{\gamma\gamma}$ within $(0.110, 0.155)$ GeV/c$^2$.

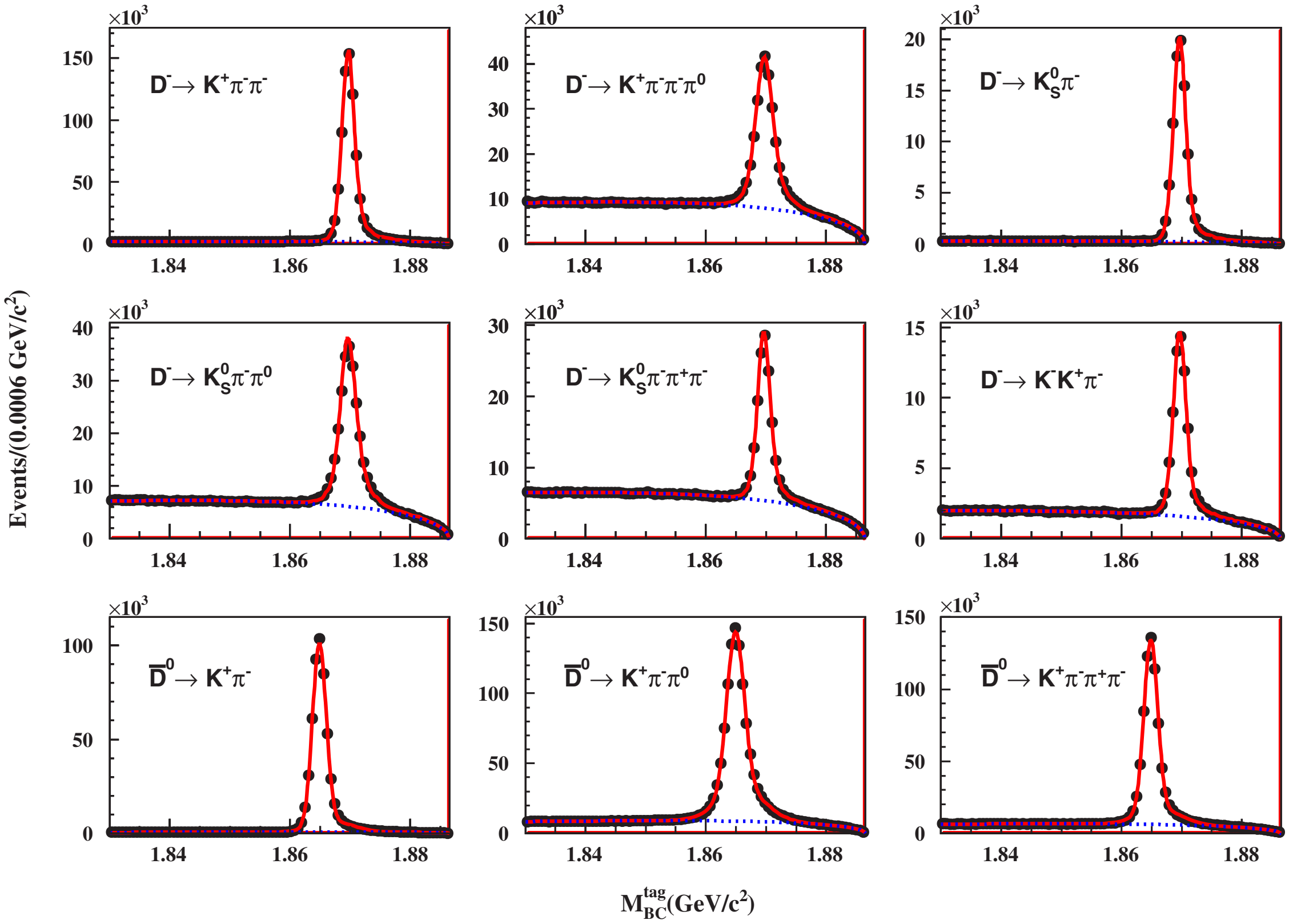

FIG. 1. Distributions of $M_{\rm BC}^{\rm tag}$ for all ST modes. Data are shown as points with error bars. The solid lines are the total fits, and the dashed lines are the background contribution.

A kinematic fit constraining $M_{\gamma\gamma}$ to the nominal $\pi^0$ mass is performed. The candidate with the smallest $\chi^2$ is kept and is required to satisfy $\chi^2 < 20$. An $\omega$ candidate is reconstructed with its $\pi^+\pi^-\pi^0$ decay mode, by requiring the three-pion invariant mass $M_{\pi^+\pi^-\pi^0}$ to be within $(0.720, 0.840)$ GeV/c$^2$. For the $K_S^0$ candidates, in addition to the same criteria as used in ST event selection, we further require $L/\sigma_L > 2$, where $L$ is the measured $K_S^0$ flight distance and $\sigma_L$ is the corresponding uncertainty.

Similar to the ST selection, $\Delta E$ and $M_{\rm BC}$ for the signal candidates of the rare $D$ decays in DT events, denoted as $\Delta E_{\rm sig}$ and $M_{\rm BC}^{\rm sig}$, are calculated. For each signal mode, $\Delta E_{\rm sig}$ is required to be within $3\sigma$ of the nominal value, as listed in Table II, and only the combination with the smallest $|\Delta E_{\rm sig}|$ is kept. The $M_{\rm BC}^{\rm sig}$ distributions of the surviving events are shown in Fig. 2, where no significant excess over the expected backgrounds is observed. The number of remaining signal candidates, $n_{\rm obs}$, is counted in the $M_{\rm BC}^{\rm sig}$ signal regions and listed in Table II. The corresponding DT detection efficiencies and the average signal efficiencies $\varepsilon_{\rm sig}$ over different ST modes are given in Table III. The BFs of the rare decays will be determined by subtracting the background contributions.

TABLE II. The $\Delta E_{\rm sig}$ requirements, the $M_{\rm BC}^{\rm sig}$ signal regions, the observed number of signal events $n_{\rm obs}$, and the estimated background yields $n_{\rm bkg1}^{\rm SB}$ and $n_{\rm bkg2}^{\rm MC} \pm \sigma_{\rm bkg2}^{\rm MC}$ in the $D^+$ and $D^0$ signal modes.

| $D^+$ decays | $\Delta E_{\rm sig}$ (GeV) | $M_{\rm BC}^{\rm sig}$ (GeV/c$^2$) | $n_{\rm obs}$ | $n_{\rm bkg1}^{\rm SB}$ | $n_{\rm bkg2}^{\rm MC} \pm \sigma_{\rm bkg2}^{\rm MC}$ |
|---|---|---|---|---|---|
| $\pi^+\pi^0e^+e^-$ | $(-0.060, 0.030)$ | (1.864, 1.877) | 4 | 0 | $5.3 \pm 0.7$ |
| $K^+\pi^0e^+e^-$ | $(-0.063, 0.037)$ | (1.862, 1.877) | 1 | 0 | $0.5 \pm 0.2$ |
| $K_S^0\pi^+e^+e^-$ | $(-0.038, 0.020)$ | (1.865, 1.877) | 6 | 0 | $4.6 \pm 0.7$ |
| $K_S^0K^+e^+e^-$ | $(-0.038, 0.021)$ | (1.865, 1.875) | 0 | 0 | $0.2 \pm 0.1$ |
| $D^0$ decays | $\Delta E_{\rm sig}$ (GeV) | $M_{\rm BC}^{\rm sig}$ (GeV/c$^2$) | $n_{\rm obs}$ | $n_{\rm bkg1}^{\rm SB}$ | $n_{\rm bkg2}^{\rm MC} \pm \sigma_{\rm bkg2}^{\rm MC}$ |
| $K^-K^+e^+e^-$ | $(-0.044, 0.015)$ | (1.858, 1.872) | 2 | 0 | $0.9 \pm 0.3$ |
| $\pi^+\pi^-e^+e^-$ | $(-0.053, 0.020)$ | (1.857, 1.873) | 11 | 2 | $11.8 \pm 1.1$ |
| $K^-\pi^+e^+e^-$ | $(-0.040, 0.018)$ | (1.857, 1.873) | 49 | 1 | $32.4 \pm 1.7$ |
| $\pi^0e^+e^-$ | $(-0.043, 0.020)$ | (1.853, 1.879) | 2 | 0 | $2.1 \pm 0.4$ |
| $\eta e^+e^-$ | $(-0.094, 0.031)$ | (1.854, 1.878) | 0 | 0 | $0.6 \pm 0.3$ |
| $\omega e^+e^-$ | $(-0.086, 0.035)$ | (1.854, 1,878) | 2 | 0 | $4.0 \pm 0.6$ |
| $K_S^0e^+e^-$ | $(-0.078, 0.035)$ | (1.858, 1.873) | 4 | 0 | $2.2 \pm 0.5$ |

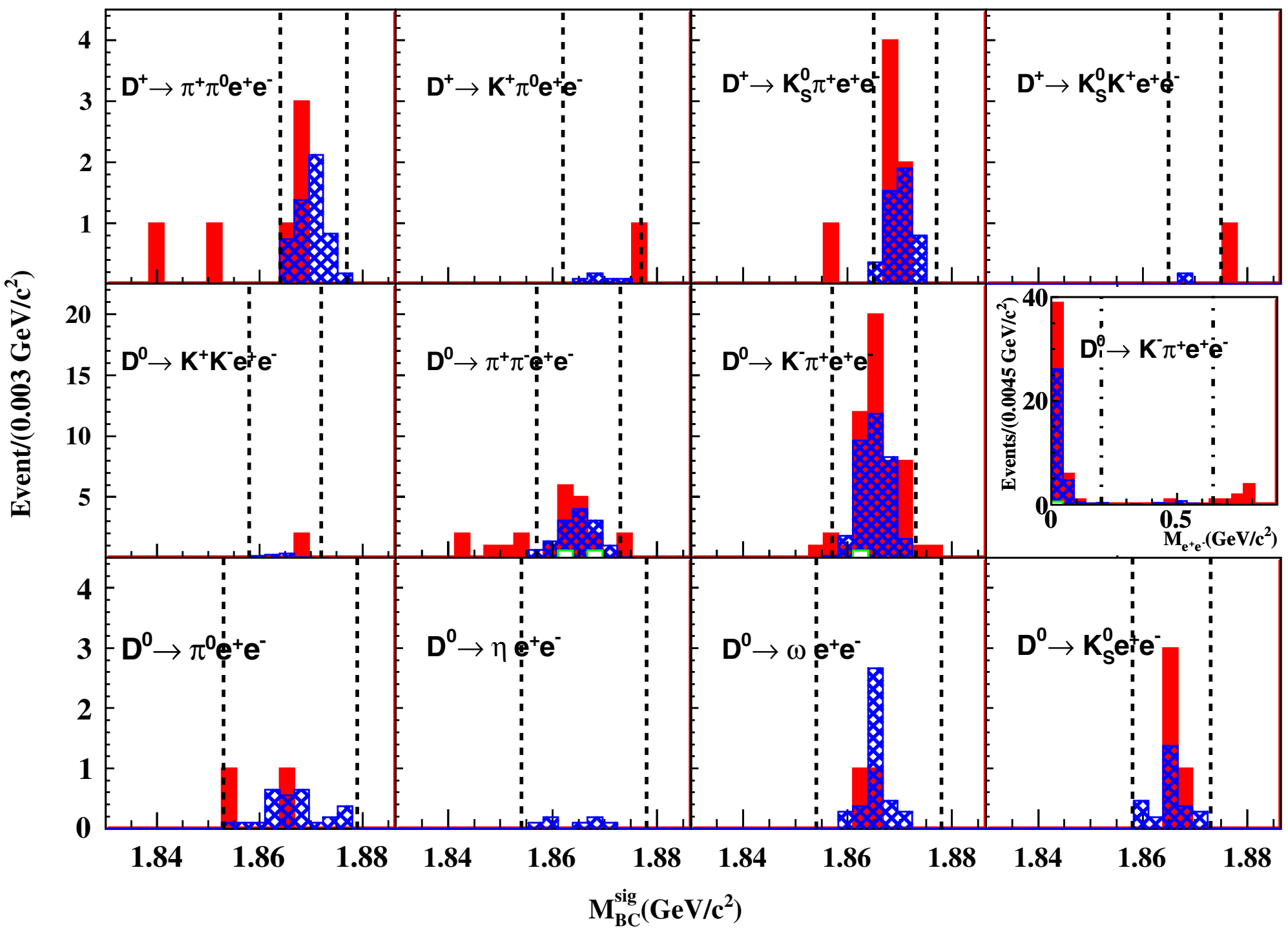

FIG. 2. Distributions of $M_{\rm BC}^{\rm sig}$ for the signal modes after applying all selection criteria. The solid histograms are data, the hatched ones are the events in the inclusive MC samples scaled to the luminosity of data, the hollow ones are the SB events in the ST $M_{\rm BC}^{\rm tag}$ distributions, and the dashed lines denote the signal regions. The inset shows the $M_{e^+e^-}$ distribution for $D^0 \to K^-\pi^+e^+e^-$, which is divided into three regions, $[0.00, 0.20)$, $[0.20, 0.65)$ and $[0.65, 0.90]$ GeV/c$^2$, distinguished by the dot-dashed lines.

The backgrounds are separated into two categories: events with a wrong ST candidate, and events with a correct ST but wrong signal candidate, which dominantly originate from the $\gamma$-conversion process. The former background can be estimated with the surviving events in the ST sideband (SB) region of $M_{\rm BC}^{\rm tag}$ distribution, which is defined as $(1.830, 1.855)$ GeV/c$^2$ for $\bar{D}^0$ decays and $(1.830, 1.860)$ GeV/c$^2$ for $D^-$ decays. The corresponding number of wrong-ST background events, $n_{\rm bkg1}$, is estimated with the number of events in the SB region ($n_{\rm bkg1}^{\rm SB}$) normalized by a scale factor $f$, which is the ratio of the integrated numbers of background events in the signal and SB regions. The scale factor $f$ is found to be $0.466 \pm 0.001$ for the $D^+$ decays and $0.611 \pm 0.001$ for the $D^0$ decays, respectively, where the uncertainty is statistical only. The wrong-ST background is expected to follow a Poisson ($\mathcal{P}$) distribution with central value of $n_{\rm bkg1} \cdot f$. The background from misreconstructed signal is estimated with the $D^+D^-$ and $D^0\bar{D}^0$ events in the inclusive MC samples by subtracting the wrong ST events, and the corresponding number of events is expected to follow a Gaussian distribution ($\mathcal{G}$), with central value $n_{\rm bkg2}^{\rm MC}$ and standard deviation $\sigma_{\rm bkg2}^{\rm MC}$. The relevant numbers are summarized in Table II.

## IV. SYSTEMATIC UNCERTAINTIES

With the DT technique, the systematic uncertainties in the BF measurements due to the detection and reconstruction of the ST $\bar{D}$ mesons mostly cancel, as shown in Eq. (1). For the signal side, the following sources of systematic uncertainties, as summarized in Table IV, are considered. All of these contributions are added in quadrature to obtain the total systematic uncertainties.

The uncertainties of tracking and PID efficiencies for $K^\pm$ and $\pi^\pm$ are studied with control samples of $D\bar{D}$ favored hadronic modes [22]. We assign an uncertainty of 1.0% per track for the tracking and 0.5% for the PID uncertainties. The tracking and PID efficiency for $e^\pm$ detection is studied using radiative Bhabha events, and the corresponding systematic uncertainty is evaluated by weighting according to the $\cos\theta$ and transverse momentum distributions of the $e^\pm$ tracks. The uncertainties for $\pi^0$, $\eta$ and $K_S^0$ reconstructions are studied with control samples of $D\bar{D}$ events. An uncertainty of 2.0% is assigned for each $\pi^0$, 1.5% for $K^0_s$, and 1.2% for $\eta$.

TABLE III. MC-determined DT detection efficiencies and the average signal efficiencies over different ST modes of the $D^+$ and $D^0$ decay modes (%). The uncertainties are all statistical.

| $\varepsilon^{\rm MC}_{\rm sig,tag}$ $(D^+)$ | $\pi^+\pi^0e^+e^-$ | $K^+\pi^0e^+e^-$ | $K^0_S\pi^+e^+e^-$ | $K^0_SK^+e^+e^-$ |
|---|---|---|---|---|
| $K^+\pi^-\pi^+$ | $10.89 \pm 0.10$ | $9.07 \pm 0.10$ | $9.38 \pm 0.10$ | $7.93 \pm 0.09$ |
| $K^+\pi^-\pi^+\pi^0$ | $4.10 \pm 0.06$ | $2.88 \pm 0.05$ | $3.37 \pm 0.06$ | $2.75 \pm 0.05$ |
| $K^0_S\pi^-$ | $11.87 \pm 0.11$ | $9.99 \pm 0.10$ | $10.17 \pm 0.10$ | $8.44 \pm 0.09$ |
| $K^0_S\pi^-\pi^0$ | $4.70 \pm 0.07$ | $3.76 \pm 0.06$ | $3.80 \pm 0.06$ | $3.18 \pm 0.06$ |
| $K^0_S\pi^-\pi^+\pi^-$ | $6.00 \pm 0.08$ | $5.08 \pm 0.07$ | $4.00 \pm 0.06$ | $3.45 \pm 0.06$ |
| $K^-K^+\pi^-$ | $8.49 \pm 0.09$ | $7.23 \pm 0.09$ | $7.42 \pm 0.09$ | $6.21 \pm 0.08$ |
| $\varepsilon_{\rm sig}$ | $19.93 \pm 0.12$ | $16.23 \pm 0.11$ | $16.65 \pm 0.11$ | $13.99 \pm 0.10$ |

| $\varepsilon^{\rm MC}_{\rm sig,tag}$ $(D^0)$ | $K^-K^+e^+e^-$ | $\pi^+\pi^-e^+e^-$ | $K^-\pi^+e^+e^-$ |
|---|---|---|---|
| $K^+\pi^-$ | $13.03 \pm 0.19$ | $25.13 \pm 0.23$ | $19.26 \pm 0.21$ |
| $K^+\pi^-\pi^+$ | $6.68 \pm 0.08$ | $12.91 \pm 0.09$ | $10.05 \pm 0.08$ |
| $K^+\pi^-\pi^+\pi^0$ | $6.52 \pm 0.10$ | $13.01 \pm 0.12$ | $9.74 \pm 0.11$ |
| $\varepsilon_{\rm sig}$ | $19.05 \pm 0.15$ | $37.14 \pm 0.18$ | $28.49 \pm 0.16$ |

| $\varepsilon^{\rm MC}_{\rm sig,tag}$ $(D^0)$ | $\pi^0e^+e^-$ | $\eta e^+e^-$ | $\omega e^+e^-$ | $K^0_Se^+e^-$ |
|---|---|---|---|---|
| $K^+\pi^-$ | $25.39 \pm 0.19$ | $23.35 \pm 0.25$ | $11.81 \pm 0.24$ | $15.47 \pm 0.25$ |
| $K^+\pi^-\pi^+$ | $13.10 \pm 0.08$ | $11.89 \pm 0.10$ | $6.74 \pm 0.10$ | $7.78 \pm 0.10$ |
| $K^+\pi^-\pi^+\pi^0$ | $13.40 \pm 0.19$ | $12.00 \pm 0.13$ | $5.69 \pm 0.13$ | $8.00 \pm 0.14$ |
| $\varepsilon_{\rm sig}$ | $37.81 \pm 0.20$ | $34.29 \pm 0.19$ | $18.01 \pm 0.19$ | $22.63 \pm 0.20$ |

The $\gamma$-conversion background is suppressed by a requirement on the distance from the reconstructed vertex of the $e^+e^-$ pair to the IP. The uncertainty due to this requirement is studied using a sample of $J/\psi \to \pi^+\pi^-\pi^0$ with $\pi^0 \to \gamma e^+e^-$ [21]. The relative difference of the efficiency between data and MC simulation is 1.8%, and is assigned as the uncertainty.

The estimated signal detection efficiencies depend on the MC simulations, which is assumed to be distributed uniformly in momentum phase space. However, theoretically there are nontrivial contributions from LD effects [4]. Alternative MC samples with LD models, in which the $e^+e^-$ pairs originate from vector mesons are also generated to estimate the signal detection efficiencies. The resultant changes on the detection efficiencies are assigned as the systematic uncertainty. The BF uncertainty for the intermediate states decays of the neutral mesons, $\mathcal{B}_{\rm inter}$, are assigned according to the world average values [9].

## V. THE UPPER LIMITS ON BRANCHING FRACTIONS

To calculate the ULs on the BFs for the signal decays, we use a maximum likelihood estimator, extended from the profile likelihood method [23]. For the detection efficiency, we assume it follows a Gaussian distribution, whose mean and width are MC-determined efficiency $\varepsilon^{\rm MC}_{\rm sig}$ and its absolute systematic uncertainty $\varepsilon^{\rm MC}_{\rm sig} \cdot \sigma^{\rm MC}_{\varepsilon}$, respectively, and $\sigma^{\rm MC}_{\varepsilon}$ includes the relative statistical and systematic uncertainties as given in Table III and Table IV. So the joint likelihood is

$$\mathcal{L} = \mathcal{P}(n_{\rm obs}, n_{\rm tag}\cdot\mathcal{B}\cdot\varepsilon_{\rm sig} + n_{\rm bkg1} + n_{\rm bkg2}) \cdot \mathcal{G}(\varepsilon_{\rm sig}, \varepsilon_{\rm sig}^{\rm MC}, \varepsilon_{\rm sig}^{\rm MC}\cdot\sigma_{\varepsilon}^{\rm MC}) \cdot \mathcal{P}(n_{\rm bkg1}^{\rm SB}, n_{\rm bkg1}\cdot f)\cdot\mathcal{G}(n_{\rm bkg2}, n_{\rm bkg2}^{\rm MC}, \sigma_{\rm bkg2}^{\rm MC}). \quad (2)$$

TABLE IV. Relative systematic uncertainties on the BFs in percent.

| Source (%) | $K^-K^+e^+e^-$ | $\pi^+\pi^-e^+e^-$ | $K^-\pi^+e^+e^-$ | $\pi^0e^+e^-$ | $\eta e^+e^-$ | $\omega e^+e^-$ | $K^0_Se^+e^-$ | $\pi^+\pi^0e^+e^-$ | $K^+\pi^0e^+e^-$ | $K^0_S\pi^+e^+e^-$ | $K^0_SK^+e^+e^-$ |
|---|---|---|---|---|---|---|---|---|---|---|---|
| $K^\pm/\pi^\pm$ tracking | 2.0 | 2.0 | 2.0 | ... | ... | 2.0 | ... | 1.0 | 1.0 | 1.0 | 1.0 |
| $K^\pm/\pi^\pm$ PID | 1.0 | 1.0 | 1.0 | ... | ... | 1.0 | ... | 0.5 | 0.5 | 0.5 | 0.5 |
| $e^\pm$ | 6.9 | 2.8 | 4.5 | 0.8 | 1.8 | 3.6 | 2.0 | 3.9 | 5.2 | 5.2 | 6.7 |
| $K^0_S$ | ... | ... | ... | ... | ... | ... | 1.5 | ... | ... | 1.5 | 1.5 |
| $\pi^0$ | ... | ... | ... | 2.0 | ... | 2.0 | ... | 2.0 | 2.0 | ... | ... |
| $\eta$ | ... | ... | ... | ... | 1.2 | ... | ... | ... | ... | ... | ... |
| $\gamma$-conversion veto | 1.8 | 1.8 | 1.8 | 1.8 | 1.8 | 1.8 | 1.8 | 1.8 | 1.8 | 1.8 | 1.8 |
| MC modeling | 3.0 | 12.8 | 24.6 | 12.6 | 13.6 | 13.1 | 6.6 | 2.1 | 5.6 | 4.9 | 4.9 |
| $B_{\rm inter}$ | ... | ... | ... | 0.1 | 0.5 | 0.8 | 0.1 | 0.1 | 0.1 | 0.1 | 0.1 |
| Total | 8.0 | 13.4 | 25.2 | 12.9 | 13.9 | 14.0 | 7.3 | 5.3 | 8.2 | 7.5 | 8.6 |

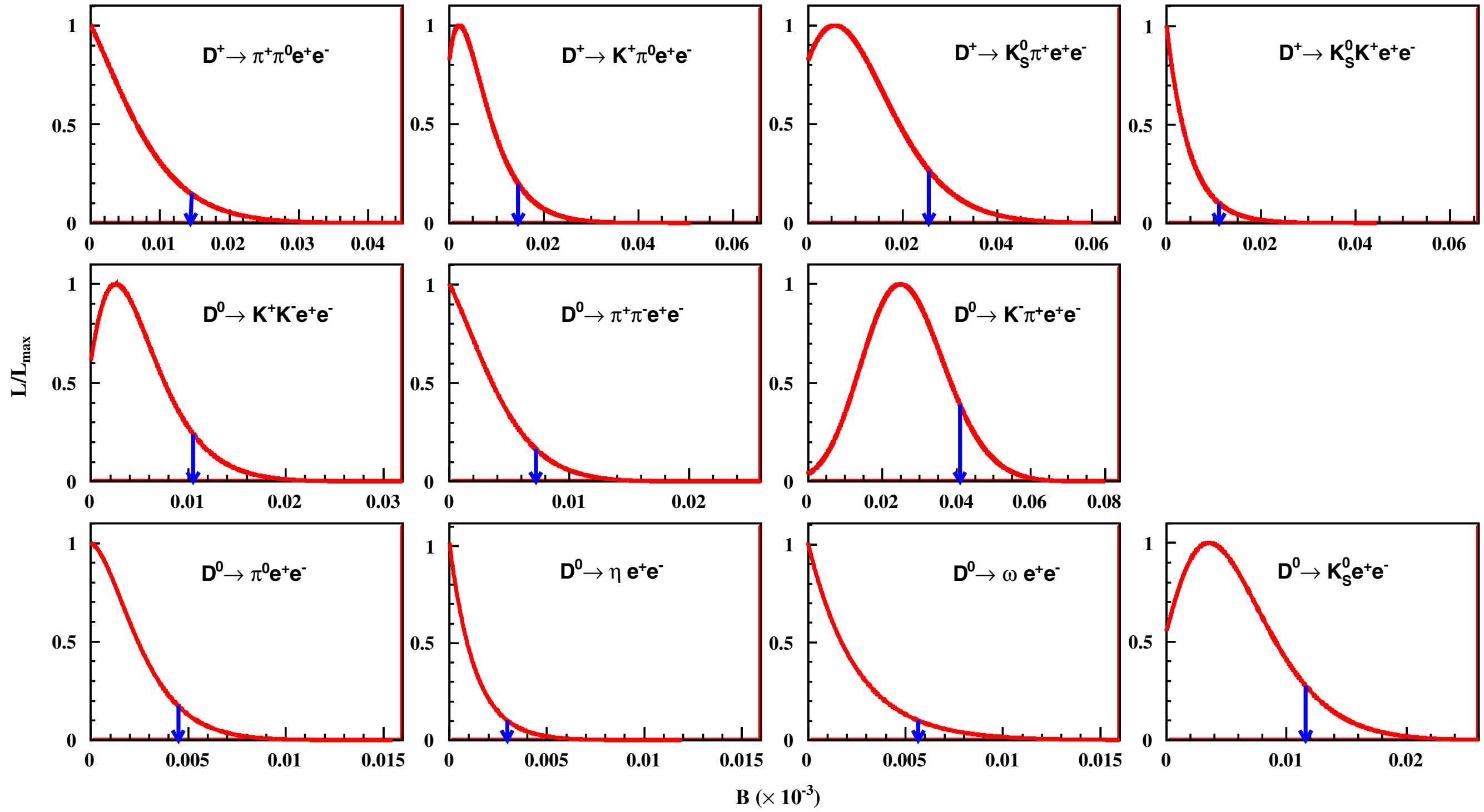

FIG. 3. Likelihood curves as a function of the signal BFs. The arrows point to the position of the ULs at the 90% CL.

TABLE V. Results of the ULs on the BFs for the investigated rare decays at the 90% CL, and the corresponding results in the PDG. Also listed are the results of the BFs in the different $M_{e^+e^-}$ regions for $D^0\to K^-\pi^+e^+e^-$. The uncertainties include both statistical and systematic ones.

| Signal decays | $\mathcal{B}$ ($\times10^{-5}$) | PDG [9] ($\times10^{-5}$) |
|---|---|---|
| $D^+\to\pi^+\pi^0e^+e^-$ | $<1.4$ | … |
| $D^+\to K^+\pi^0e^+e^-$ | $<1.5$ | … |
| $D^+\to K_S^0\pi^+e^+e^-$ | $<2.6$ | … |
| $D^+\to K_S^0K^+e^+e^-$ | $<1.1$ | … |
| $D^0\to K^-K^+e^+e^-$ | $<1.1$ | $<31.5$ |
| $D^0\to\pi^+\pi^-e^+e^-$ | $<0.7$ | $<37.3$ |
| $D^0\to K^-\pi^+e^+e^{-\dagger}$ | $<4.1$ | $<38.5$ |
| $D^0\to\pi^0e^+e^-$ | $<0.4$ | $<4.5$ |
| $D^0\to\eta e^+e^-$ | $<0.3$ | $<11$ |
| $D^0\to\omega e^+e^-$ | $<0.6$ | $<18$ |
| $D^0\to K_S^0e^+e^-$ | $<1.2$ | $<11$ |
| $^\dagger$ in $M_{e^+e^-}$ regions: | | |
| $[0.00, 0.20)$ GeV/c$^2$ | $<3.0$ $(1.5^{+1.0}_{-0.9})$ | … |
| $[0.20, 0.65)$ GeV/c$^2$ | $<0.7$ | … |
| $[0.65, 0.90)$ GeV/c$^2$ | $<1.9$ $(1.0^{+0.5}_{-0.4})$ | … |

Based on the Bayesian method, we use the likelihood distribution as a function of the signal BF $\mathcal{B}$, with variations of the other parameters $n_{\rm bkg1}$, $n_{\rm bkg2}$, and $\varepsilon_{\rm sig}$, as the probability function. Note that the ST yields, $n_{\rm tag}$, are taken as the truth ones, as their uncertainties are negligible.

The resultant likelihood distributions for all the signal modes are shown in Fig. 3, and the ULs on the signal BFs at the 90% CL are estimated by integrating the likelihood curves in the physical region of $\mathcal{B}\geq 0$. For $D^0\to K^-\pi^+e^+e^-$, the BF is determined to be $(2.5\pm1.1)\times10^{-5}$ with a significance of $2.6\sigma$, where the uncertainty includes the statistical and systematic ones. Reference [4] predicts the BF of $D^0\to K^-\pi^+e^+e^-$, which is dominated by the LD bremsstrahlung and (virtual) resonance-decay contributions in the lower and upper regions, respectively, to exceed $0.99\times10^{-5}$ in the lower $M_{e^+e^-}$ region, adding up to $1.6\times10^{-5}$ in the whole region. Therefore, we divide the $M_{e^+e^-}$ distribution into three regions and determine the BFs in the individual regions. All these results are listed in Table V, and are all within the SM predictions.

## VI. SUMMARY

To summarize, searches for $D^+$ and $D^0$ decays into $h(h^{(\prime)})e^+e^-$ final states are performed, based on the DT analysis of a $e^+e^-$ collision sample of 2.93 fb$^{-1}$ taken at $\sqrt{s}=3.773$ GeV with the BESIII detector. No evident signals are observed, and the corresponding ULs on the decay rates are determined at the 90% CL, as shown in Table V. For the four-body $D^+$ decays, the searches are performed for the first time. The reported ULs of the $D^0$ decays are improved in general by a factor of 10, compared to previous measurements [9]. All the measured ULs on the BFs are above the SM predictions [3,4], which include both LD and SD contributions.

## ACKNOWLEDGMENTS

The BESIII collaboration thanks the staff of BEPCII and the IHEP computing center for their strong support. This work is supported in part by National Key Basic Research Program of China under Contract No. 2015CB856700; National Natural Science Foundation of China (NSFC) under Contracts No. 11235011, No. 11275266, No. 11335008, No. 11425524, No. 11625523, No. 11635010; the Chinese Academy of Sciences (CAS) Large-Scale Scientific Facility Program; the CAS Center for Excellence in Particle Physics (CCEPP); Joint Large-Scale Scientific Facility Funds of the NSFC and CAS under Contracts No. U1332201, No. U1532257, No. U1532258; CAS under Contracts No. KJCX2-YW-N29, No. KJCX2-YW-N45, No. QYZDJ-SSW-SLH003; 100 Talents Program of CAS; National 1000 Talents Program of China; INPAC and Shanghai Key Laboratory for Particle Physics and Cosmology; German Research Foundation DFG under Contracts No. Collaborative Research Center CRC 1044, No. FOR 2359; Istituto Nazionale di Fisica Nucleare, Italy; Koninklijke Nederlandse Akademie van Wetenschappen (KNAW) under Contract No. 530-4CDP03; Ministry of Development of Turkey under Contract No. DPT2006K-120470; National Science and Technology fund; The Swedish Research Council; U.S. Department of Energy under Contracts No. DE-FG02-05ER41374, No. DE-SC-0010118, No. DE-SC-0010504, No. DE-SC-0012069; University of Groningen (RuG) and the Helmholtzzentrum fuer Schwerionenforschung GmbH (GSI), Darmstadt; WCU Program of National Research Foundation of Korea under Contract No. R32-2008-000-10155-0.